# Development of Non-Linear Equations for Predicting Electrical Conductivity in Silicates

Patrick dos Anjos[1]*
Lucas de Almeida Quaresma*
Marcelo Lucas Pereira Machado*

*Federal Institute of Espírito Santo (IFES), Vitória, ES, Brazil*

## Abstract

Electrical conductivity is of fundamental importance in electric arc furnaces (EAF) and the interaction of this phenomenon with the process slag results in energy losses and low optimization. As mathematical modeling helps in understanding the behavior of phenomena and it was used to predict the electrical conductivity of EAF slags through artificial neural networks. The best artificial neural network had 100 neurons in the hidden layer, with 6 predictor variables and the predicted variable, electrical conductivity. Average absolute error and standard deviation of absolute error were calculated, and sensitivity analysis was performed to correlate the effect of each predictor variable with the predicted variable.

## 1    Introduction

The electric arc furnace (EAF) slag is composed of $SiO_2$-$CaO$-$MgO$-$Al_2O_3$-$FeO$ [1,2] system with a temperature of up to 1756K [3]. The electrical conductivity of the slag in the EAF is important in the industrial process and directly affects the quality of the final product and energy consumption [4].

Slag foaming consists of introducing gas bubbles into molten metal and slag by bubble injection or chemical reaction. The slag foam protects the refractory applied to the EAF of the arc, increasing the working time of the lining composed of refractories. The foam prevents oxidation of the molten material and allows control of its chemical composition, aiding refining and homogenization, also acting as a thermal insulator between the molten material and its surroundings, thus reducing the energy required to maintain the operating temperature [2].

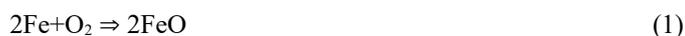
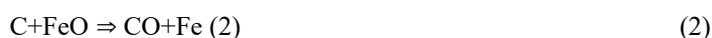

$$2Fe + O_2 \Rightarrow 2FeO \qquad (1)$$
$$C + FeO \Rightarrow CO + Fe \qquad (2)$$

---
1    E-mail: patrick.dosanjos@outlook.com

To form and maintain the CO gas bubbles responsible for slag foaming, formed through the Equation 1 in steel and in slag reactions (Equation 2) (Figure 1), optimized slag chemistry is required. The slag, also composed of solid phases, mainly formed by MgO and CaO, must have a viscosity in a narrow range that generally occurs when there is the formation of the MgO·FeO phase through the saturation of MgO [5].

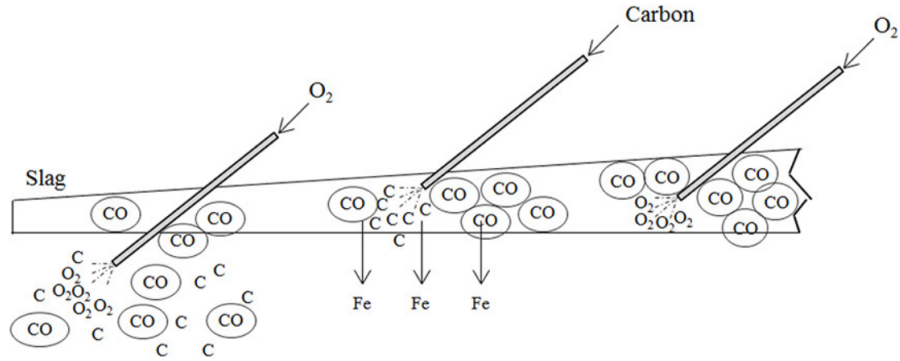

Figure 1 — The action of Fe, $O_2$ and C in the formation of CO bubbles [6].

The electrical conductivity of slag can generally be determined by electronic conduction and ionic conduction [7] correlating the structure of the slag composed by the NBO/T parameter. In $CaO-SiO_2-B_2O_3$ slags, charge transport occurs through $Ca^{2+}$ ions because $Si^{4+}$ and $B^{3+}$ ions belong to the structure of this silicate [4]. Generally, ions with high valence number interact with the surrounding ions forming structural parts of the slag and thus low mobility, not contributing to the electrical conductivity of the material [4] (Figure 2).

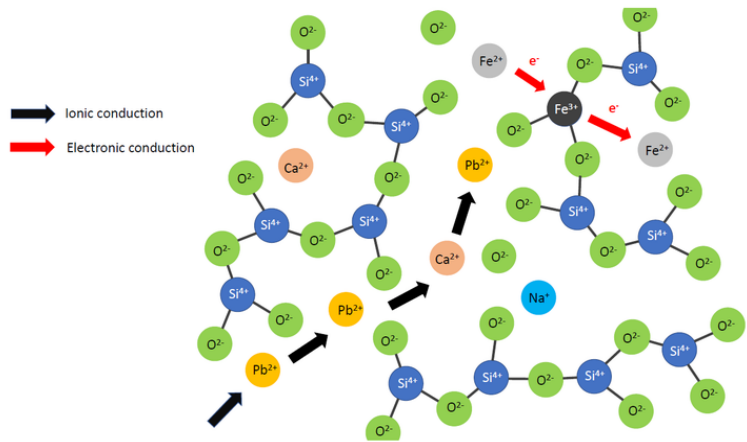

Figure 2. Electrical conductivity by electronic and ionic conduction in slags.

The modeling of electrical conductivity can be carried out by linear methods [4] but it has the drawback of presenting only linearity in the modeling process. With that, the non-linear modeling is able to capture the complexity of the process of mathematical modeling of the electrical conductivity of slag. With the use of artificial neural networks, non-linear modeling was performed with optimization by hyperparameter variation. The present work aims at the non-linear modeling of the electrical conductivity of $SiO_2-CaO-MgO-Al_2O_3-FeO$ slags using artificial neural networks.

## 2 Materials and Methods

### 2.1 Database

The Sciglass database used to predict the refractive index of optical materials [8] and the viscosity of oxides [9] was chosen to provide data on electrical conductivity (Siemens/m or S/m), temperature (K) and chemical composition of $SiO_2$-CaO-MgO-$Al_2O_3$-FeO (molar fraction) silicates.

The electrical conductivity histogram can be seen in Figure 3.

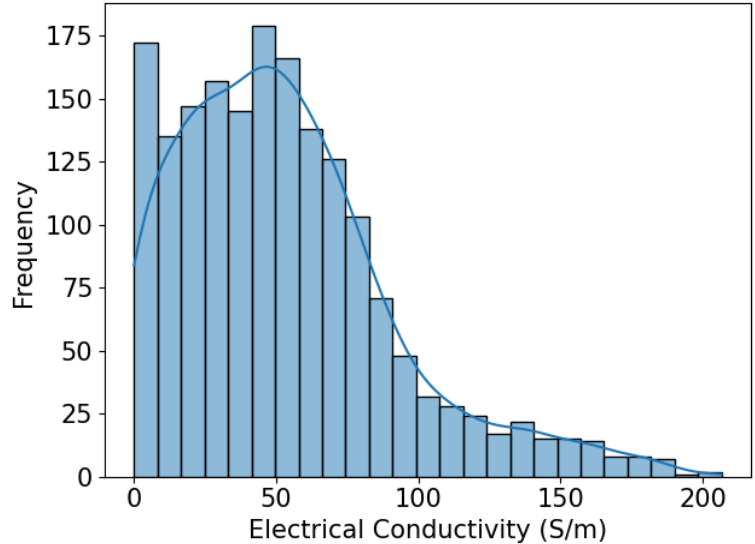

Figure 3. Electrical conductivity histogram.

All data were submitted in the removal of outliers where data above or below 3 standard deviations of electrical conductivity is removed. An optimization method during the training of artificial neural networks stems from changing the range of input variables to compute the algorithm with greater efficiency and speed. The standardization method [9] uses calculus to result in a normal distribution for each input variable in the artificial neural network, with the downside of modifying the variable's distribution. The normalization between values makes it possible to enter with the same distribution as the data prior to preprocessing and improves training time.

Artificial neural networks are algorithms that try to achieve performance with respect to learning from experience, and thereby make generalizations from similar situations and judge states where good and bad results were achieved in the past [10]. The statistical pattern approach has been the most commonly studied and used in practice in the implementation and construction of artificial neural networks, which is also referred to as a non-linear model that replicates the biological neuron [11].

With extensive parameters [9], such as the number of neurons, layers, activation functions, optimizers, weights and bias initialization, loss and metrics, the construction and training along with the testing phase of an artificial neural network can be time-consuming and difficult. But artificial neural networks have advantages over other mathematical modeling practices because they are robust and develop complex relationships between variables [12], in addition to presenting a modeling that has the properties of a universal approximator.

## 2.2 Artificial Neural Networks

Artificial neural networks (ANN) with a fixed number of depth and an arbitrary number of neurons can approximate any continuous functions in a closed set with an inherent error when the activation function applied in the construction of this artificial neural network is continuous and non-polynomial [13]. This therefore indicates that artificial neural networks with a fixed depth of 1 with a variable number of neurons is also a universal approximator. There is also a minimum width for artificial neural networks to present the properties of a universal approximator as a function of the cardinality of the set of arguments in the mathematical modeling [13].

The artificial neural networks were built with the variation of neurons following the equation where the minimum number of neurons in the hidden layer ($W_{min}$) must be equal to or greater than the cardinality of the set of arguments of the artificial neural network ($D_x$) added to 1 (Equation 3) to provide the universal approximator property [14].

$$W_{min} \geq D_x + 1 \qquad (3)$$

The Adam [15] optimizer was used, which is an algorithm for first-order gradient-based optimization of stochastic objective functions. For the initialization of weights and biases, the method of initialization of weights and biases Glorot uniform [16] was used, which has advantages over the methods of initialization of standardized weights and biases because it presents lower activation values and backpropagated gradients with approximation at different depths in an artificial neural network.

With the preprocessing database, training data and test data were taken to present efficiency values through the chosen metrics. The amount of 80% for the training phases and 20% for the testing steps were separated and the test data were **not** used during the training of the artificial neural networks. The test steps were carried out with a loss calculated by the root mean squared error (Equation 4) [17]. The activation function ReLU (Equation 5) was chosen as the activation function of the trained artificial neural networks.

$$RMSE = \sqrt{\frac{1}{N} \sum (y_{true} - y_{predicted})^2} \qquad (4)$$

$$f(x) = \max(0, x) \qquad (5)$$

Thus, the artificial neural networks were built by varying the neurons in the hidden layer with 6 input variables (Temperature in Kelvin and chemical composition of $SiO_2$, $CaO$, $MgO$, $Al_2O_3$, $FeO$ in molar fraction) and 1 output variable (Electrical conductivity in S/m).

## 2.3 Statistical Evaluation

Metrics of average absolute error (Equation 6) [18] and standard deviation (Equation 7) [19] were used during the training and test phase of the artificial neural networks. The standard deviation can be related to the shape of a one-variable distribution where it indicates the width of that variable in a probability density function [20].

$$AAE = \frac{1}{N} \sum |y_{true} - y_{predicted}| \qquad (6)$$

$$St.Dev. = \sqrt{\frac{1}{N} \sum (deviation - \mu_{deviation})^2} \qquad (7)$$

*deviation* = $y_{true} - y_{predicted}$ and $\mu_{deviation}$ is the arithmetic mean of deviation.

Sensitivity analysis can be defined as the determination of the contribution spectrum of an input variable in an artificial neural network [21]. There are different ways of performing sensitivity analysis in an artificial neural network, such as the perturbation of the input variables, partial derivatives [21] and the method of connection weights [22]. The connection weights method is determined by the relationships between the weights of an artificial neural network and applied to perform the sensitivity analysis to predict the density of oil-based muds in high-temperature [23] and the viscosity of multicomponent slags [24]. Furthermore, the sensitivity analysis using the connection weights method has a higher similarity coefficient than other sensitivity analysis methods [21].

The best artificial neural network was subjected to the connection weights method in relation to sensitivity analysis to demonstrate the relative importance of each input variable, chemical composition and temperature, in relation to the output variable, electrical conductivity.

## 3   Results and discussion

Several artificial neural networks were trained and the best one had a number of 100 neurons in the hidden layer. Therefore, the best artificial neural network has 6 input variables, 100 neurons in the hidden layer and 1 neuron in the output layer. The training of the best artificial neural network in relation to losses at training epochs (Figure 4) has a high loss value in the first training epochs later optimized by the Adam algorithm, presenting a loss value close to zero at the end of training.

The best artificial neural network presented a average absolute error result of 21.29 S/m and a standard deviation of 20.07 S/m (Table 1) and the graph that relates the deviation between the database data and the predicted data for the best artificial neural network can be seen in Figure 5.

Table 1. Average absolute error (AAE) and Standard Deviation in ANN

| Model | AAE (S/m) | St. Dev. (S/m) |
|---|---|---|
| ANN | 21.29 | 20.07 |

With the predicted data there is a good approximation in relation to the database raised through the use of the chemical composition data of the $SiO_2$-$CaO$-$MgO$-$Al_2O_3$-$FeO$ system and the temperature in relation to the electrical conductivity of the slag applied and generated in the EAF. Mathematical modeling also helps in decision-making in control processes based on sensitivity analysis. With the weights of the hidden layer and the output layer, the relative contributions of each input variable in relation to the output variable were calculated to establish the relative importance.

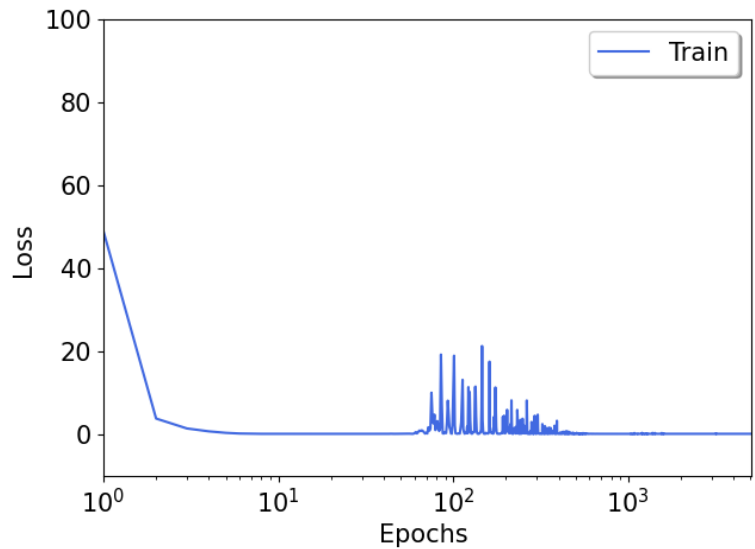

Figure 4. Loss (RMSE) versus Epochs in the best artificial neural network.

Sensitivity analysis showed that CaO and $SiO_2$ variables have a greater impact on the electrical conductivity of the studied slags, with 30.28% and 20.29% of importance, respectively. A high relative importance in relation to the CaO variable indicates that the conductivity process in the $SiO_2$-$CaO$-$MgO$-$Al_2O_3$-$FeO$ system is ionic, similar to $CaO$-$SiO_2$-$B_2O_3$ systems and the variable $SiO_2$ is a deleterious variable to the refining process of materials within an EAF.

The variables MgO and FeO showed relative importance of 19.29% and 18.67% respectively. The influence of MgO on the foaming index in EAF slags is notorious, which has extreme importance in energy efficiency in energy consumption in the process [2] and FeO acts directly in the formation of CO bubbles that act for slag foaming. $Al_2O_3$ and temperature obtained the importance of 9.22% and 2.25% respectively, demonstrating lower importance in the electrical conductivity control process in relation to the other variables (Figure 6).

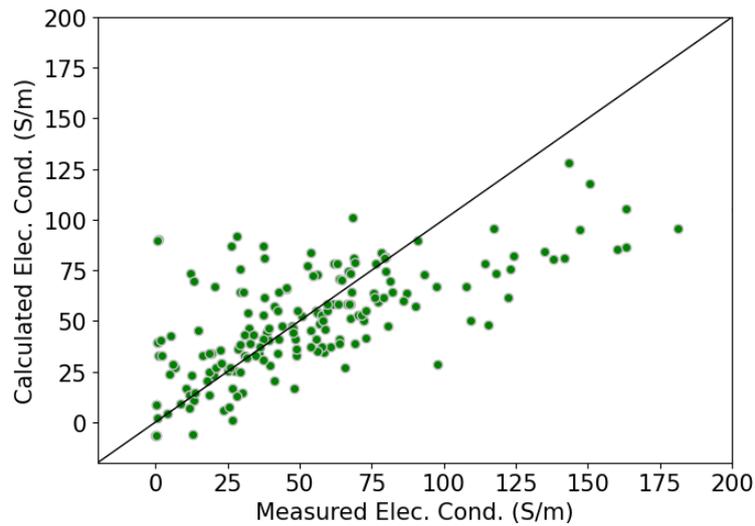

Figure 5. Electrical conductivity predicted (y-axis) and in the database (x-axis) in the developed artificial neural network.

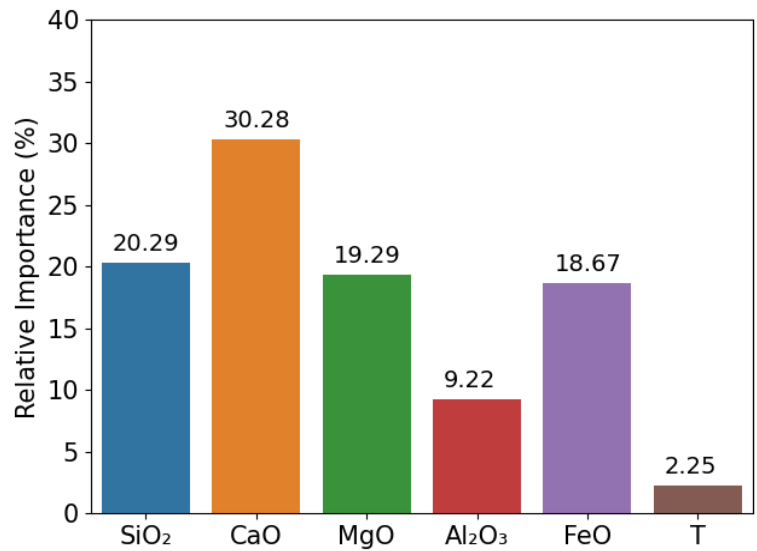

Figure 6. Relative importances in $SiO_2$-$CaO$-$MgO$-$Al_2O_3$-$FeO$ system and temperature

## 4    Conclusion

The electrical conductivity of the slag in the material manufacturing process in the electric arc furnace (EAF) has recognized importance for the quality of the final products and for the productive energy control. The use of mathematical modeling helps in understanding the interactions between some important variables, such as the chemical composition of the slag and the operating temperature.

Artificial neural networks were built for the mathematical modeling of the electrical conductivity of the EAF slag using the non-linear modeling method. The best artificial neural network showed a average absolute error of 21.29 S/m and a standard deviation of 20.07 S/m in relation to the Sciglass database used.

Sensitivity analysis using the connection weights method showed that the variable with the highest relative importance was CaO with 30.28%, thus indicating that the charge transport mechanism inn $SiO_2$-$CaO$-$MgO$-$Al_2O_3$-$FeO$ system is by ionic conduction. The second highest relative importance was $SiO_2$ with 20.29%, which has a deleterious effect on refining reactions in the EAF. Then the variables MgO and FeO indicated relative importance of 19.29% and 18.67% indicating that both have similar values. The variables $Al_2O_3$ and temperature had the lowest importance relative to the 6 input variables in the best artificial neural network, presenting 9.22% and 2.25% respectively.